\documentclass[twocolumn]{el-author}
\usepackage{cite}
\usepackage{color,soul}
\makeatletter
\newcommand{\vast}{\bBigg@{4}}
\newcommand{\Vast}{\bBigg@{5}}
\makeatother
\makeatletter
\newenvironment{sqcases}{%
  \matrix@check\sqcases\env@sqcases
}{%
  \endarray\right.%
}
\def\env@sqcases{%
  \let\@ifnextchar\new@ifnextchar
  \left\lbrack
  \def\arraystretch{1.2}%
  \array{@{}l@{\quad}l@{}}%
}
\makeatother

\makeatletter
\newlength \figwidth
\if@twocolumn
\else
\fi
\makeatother

\newcommand{\ua}{\uparrow}
\newcommand{\nc}{\newcommand}
\nc{\da}{\downarrow} \nc{\hc}{\hat{c}} \nc{\hS}{\hat{S}}
\nc{\bra}{\langle} \nc{\ket}{\rangle} \nc{\eq}{equation (\ref}
\nc{\h}{\hat} \nc{\hT}{\h{T}}\nc{\be}{\begin{eqnarray}}
\nc{\ee}{\end{eqnarray}}\nc{\rd}{\textrm{d}}\nc{\e}{eqnarray}\nc{\hR}{\hat{R}}\nc{\Tr}{\mathrm{Tr}}
\nc{\tS}{\tilde{S}}\nc{\tr}{\mathrm{tr}}\nc{\8}{\infty}\nc{\lgs}{\bra\ua,\phi|}\nc{\rgs}{|\ua,\phi\ket}
\nc{\hU}{\hat{U}}\nc{\lfs}{\bra\phi|}\nc{\rfs}{|\phi\ket}\nc{\hZ}{\hat{Z}}\nc{\hd}{\hat{d}}\nc{\mD}{\mathcal{D}}
\nc{\bd}{\bar{d}}\nc{\bc}{\bar{c}}\nc{\mc}{\mathcal}\nc{\ea}{eqnarray}\nc{\mG}{\mathcal{G}}\nc{\bce}{\begin{center}}
\nc{\ece}{\end{center}}
\date{}

\begin{document}

\title{Cooperative communications for sleep monitoring in wireless body area networks}

\author{Samiya M. Shimly$^*$, Samaneh Movassaghi, David B. Smith}

\abstract{This paper investigates the performance of cooperative receive diversity, for the wireless body area network (WBAN) radio channel, compliant with the IEEE $802.15.6$ Standard, in the case of monitoring a sleeping person. Extensive WBAN measurements near the $2.4$ GHz ISM band were used. Up to $7$ dB and $20$\% improvement for two-hop communications with the use of relays are empirically demonstrated with respect to outage probability and outage duration, with $3$-branch cooperative selection combining and $3$-branch cooperative switch-and-examine combining.}

\maketitle

\section{Introduction}

Wireless Body Area Networks (WBANs) are low power, short-range communications networks of sensors/actuators placed on, in, around and/or near the human body. One of the principle application areas of WBANs is likely to be health-care \cite{1}. Sleeping-body monitoring is important, as the patient using WBANs in hospitals, aged-care facilities, rehabilitation units or home health-care, spends a significant amount of time sleeping, which requires reliable operation of the WBAN.
  Experimental studies of propagation characteristics for the particular WBAN sleeping channel for on-body and off-body transmissions have been investigated in \cite{2}. Also, the authors in \cite{2} proposed possible use of relays to overcome unreliable communications. As cooperative diversity has great advantages in resource constraint networks like WBANs \cite{3}, it is important to determine possible performance improvements in WBANs by using relays, with different diversity combining techniques.
  
 \section{Experimental Setup}

Extensive on-body and off-body channel gain data taken over at least $2$ hours per measurement set from eight adult sleeping subjects were used for testing the relayed cooperative scheme. Each measurement set was implemented using $7$ small wearable radios operating at $2.36$ GHz, of which three were transceivers ($\rm{T_x}$/$\rm{R_x}$) and four were receivers ($\rm{R_x}$). The wearable radios are described in \cite{4} and the experimental datasets are available for download in \cite{5}. Radios were placed on different body-parts of the sleeping subject in bed. Some of them were also placed around the bed. The positions are given in Table $1$. 
Every transmitter was transmitting for $5$ms in round-robin fashion. Hence, the received signal strength indicator (RSSI), in dBm, was captured from any given transmitter every $15$ms. With a $\rm{T_x}$ power of $0$ dBm, each RSSI measurement provides an equivalent channel gain (magnitude) in dB for each $\rm{T_x}$ to $\rm{R_x}$ packet transmission.

\begin{table}[h]
\processtable{Radio positions. $\rm{NTB_h}$: Next To Bed (head), $\rm{L_w}$: Left wrist, $\rm{H_f}$: Hip front, $\rm{R_w}$: Right wrist, $\rm{H_b}$: Hip back, $\rm{L_a}$: Left ankle, $\rm{NTB_f}$: Next To Bed (foot); $\rm{T_x}$/$\rm{R_x}$ implies to Transceiver and $\rm{R_x}$ implies to Receiver}
{\begin{tabular}{ |l|l|l|l|l|l|l| }\hline
$\rm{NTB_h}$ & $\rm{L_w}$ & $\rm{H_f}$ & $\rm{R_w}$ & $\rm{H_b}$ & $\rm{L_a}$ & $\rm{NTB_f}$\\\hline
$\rm{T_x}$/$\rm{R_x}$ & $\rm{T_x}$/$\rm{R_x}$ & $\rm{T_x}$/$\rm{R_x}$ & $\rm{R_x}$ & $\rm{R_x}$ & $\rm{R_x}$ & $\rm{R_x}$\\\hline
\end{tabular}}{}
\end{table}

\begin{figure}[h]
\centering{\includegraphics[width=\figwidth]{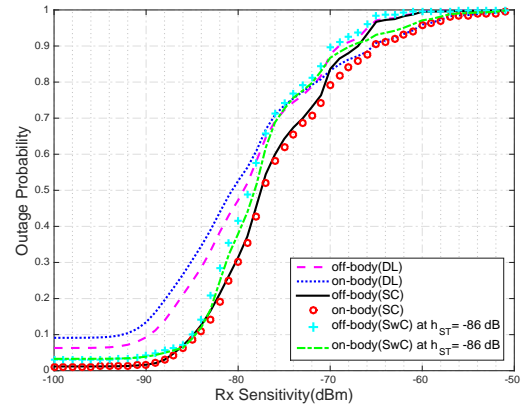}}
\caption{Outage probability as a function of receive sensitivity, with $T_x$ power of 0 dBm, for direct link (DL), switch-and-examine combining (SwC) and selection combining (SC)}
\end{figure}

\section{System Model}
WBAN channels are suitable for investigating cooperative receive diversity using relays because of their stability and reciprocity. The channels are considered to be stable, since 15ms is significantly less than the channel coherence time for a WBAN, even in a highly dynamic scenario \cite{6}. Also, due to the reciprocity property, the channel from any $\rm{T_x}$ at position $a$ to $\rm{R_x}$ at position $b$ is similar for $\rm{T_x}$ at $b$ to $\rm{R_x}$ at $a$ \cite{4}. Effectively, simultaneous measurements, based on channel coherence, describe the channel gains for relayed packets from $\rm{T_x}$-$\rm{R_{x_{relay}}}$-$\rm{R_x}$, where a direct link and two relayed links are accounted for. Three-branch cooperative selection combining (SC) and cooperative switch-and-examine combining (SwC) were investigated and compared with non-cooperative direct link (where one of the branches was a direct link and the other two were cooperative relay links). The RSSI-based narrowband channel gain (used as a negative measure of the channel attenuation) statistics were used for result estimation. For two-hop communications, where every packet at a relay is transmitted to the hub, the channel gain for one diversity branch can be described as  $h_{s{r_n}d}$ = min\{{}$h_{s{r_n}}$,  $h_{{r_n}d}$\}{}, for any given relay $r_n$ where $s$ indicates source (sensor) and $d$ indicates destination (hub). $h_{s{r_n}}$ and  $h_{{r_n}d}$ are the channel gains from source to $n^{th}$ relay and $n^{th}$ relay to destination, respectively. Thus, for three-branch cooperative selection combining, where one of the branches is a direct link, the equivalent channel gain at the output of the selection combining is shown in equation ($1$) as follows:
\begin{align}
 h_{sc} = \max\big\{h_{sd}, h_{s{r_1}d}, h_{s{r_2}d}\big\} 
\end{align}
where, $h_{sd}$  is the channel gain from source-to-destination (direct branch), $h_{s{r_1}d}$ and $h_{s{r_2}d}$ are the channel gain of first and second diversity branch, respectively. For switch-and-examine combining, three-branch switch-and-examine combining is used, where a switch to another branch occurs when the channel gain on the current branch goes below a given threshold, $h_{ST}$ at time instant $\tau$ \cite{6}. However, if the channel gain on this alternate branch is also less than $h_{ST}$; then, another switch occurs to the last-branch and, whether or not the channel gain of this branch is above or below $h_{ST}$, it becomes the chosen branch. The process is described in equation ($2$) as follows:

\begin{align}
h_{sw}(\tau)=\begin{cases} h_{sd}(\tau),\iff & (h_{sd}(\tau)\geq h_{ST}) \wedge\\
&\begin{cases}(h_{sw}(\tau-1) = h_{sd}(\tau-1))\\ \vee(h_{s{r_2}d}(\tau)<h_{ST})
\end{cases}\Bigg\}\\
h_{s{r_1}d}(\tau),\iff & (h_{s{r_1}d}(\tau)\geq h_{ST}) \wedge\\
&\begin{cases}(h_{sw}(\tau-1) = h_{s{r_1}d}(\tau-1))\\ \vee(h_{sd}(\tau)<h_{ST})
\end{cases}\Bigg\}\\
h_{s{r_2}d}(\tau),\iff & \begin{sqcases}(h_{s{r_2}d}(\tau)\geq h_{ST})\wedge\\
\begin{cases}(h_{s{r_1}d}(\tau)<h_{ST})\vee\\(h_{sw}(\tau-1)=h_{s{r_2}d}(\tau-1))
\end{cases}\Bigg\}
\end{sqcases}\vast]\\
&\vee\begin{cases}(h_{sd}(\tau)<h_{ST})\wedge\\
(h_{s{r_1}d}(\tau)<h_{ST})
\end{cases}\Bigg\}
\end{cases}
\end{align}
where, $h_{sw}(\tau)$ is the equivalent channel gain from the output of switch-and-examine combining at time instant $\tau$.
Choosing the switching threshold is important for the relative performance of switch-and-examine combining. In this paper, an optimum threshold of  $-86$ dB is used for three-branch switch-and-examine combining according to \cite{7}, due to optimal switching rate achieved at this threshold.

\section{Performance Analysis}
Because of the relatively slow varying and non-stationary nature of the channel \cite{8}, the WBAN radio channel for sleeping activity is a difficult channel to communicate over. Due to shadowing by the body parts, transmission links may be completely blocked for very long periods of time (as people move relatively little during sleeping) \cite{9}. Hence, the outage probability and the outage duration are two main factors for analyzing the performance of sleeping-body-monitoring WBANs.

For proper estimation of outages, the effect of non-recorded measurements was referred as incorrectly decoded packets and set to a value less than the radio's receive sensitivity ($\sim-100$ dBm). The outage probability was then estimated according to the mode of operation described in equation ($1$) for cooperative selection combining and equation ($2$) for cooperative switch-and-examine combining. The empirical outage probability for direct link (DL) and cooperative combining (SC \& SwC), with on-body and off-body channels are shown in Fig. $1$. At $10$\% outage probability, there is $3$ dB and $7$ dB performance improvement of cooperative combining over direct link for off-body and on-body channels, respectively. 
Also, the best-case outage probability (for both on-body and off-body channels) is almost $3$\% and $1.2$\% for cooperative switch-and-examine combining and cooperative selection combining, respectively. This implies better improvement over the best-case outage probability of direct link for on-body ($9.1$\%) and off-body ($6.3$\%) channels. Additionally, this indicates better diversity gain for $3$-branch SC over $3$-branch SwC. Although, switch-and-examine combining has significantly lower switching rate and reduced complexity than that of selection combining \cite{7}.

As the WBAN sleeping channels are very slowly varying in nature \cite{2}, the duration of outages with respect to particular receive sensitivities are of significant importance. Fig. $2$ and Fig. $3$ illustrate the percentage of time that continuous outages larger than $x$ seconds (on horizontal axis) occur with direct link and cooperatively combined links for on-body and off-body channels, correspondingly. As can be seen in Fig. $2$, an on-body receiver with a receive sensitivity of $-86$ dBm, will experience outages of larger than $10$ seconds for more than $16$\% of the time in case of direct link; whereas, the occurence has reduced to $2$\% for cooperatively combined links. As per Fig. $3$, an off-body receiver with a receive sensitivity of $-86$ dBm, will experience outages of larger than $10$ seconds for more than $11$\% of the time for direct link, while for cooperatively combined links, outages of larger than $10$ seconds will occur less than $2$\% of the total measured time.
In addition, outages of larger than a typical latency requirement of $125$ ms \cite{10,11}, occur almost $24$\% (Fig. 2) and $17$\% (Fig. 3) of the time for on-body and off-body direct links, respectively, while it has been considerably reduced to $4$\% for both on-body and off-body cooperatively combined links (Fig. $2$ \& Fig. $3$, respectively). The empirical results are shown in Table $2$, as per results obtained from Fig. $1$, Fig. $2$ and Fig. $3$.
\begin{figure}
\centering{\includegraphics[width=\figwidth]{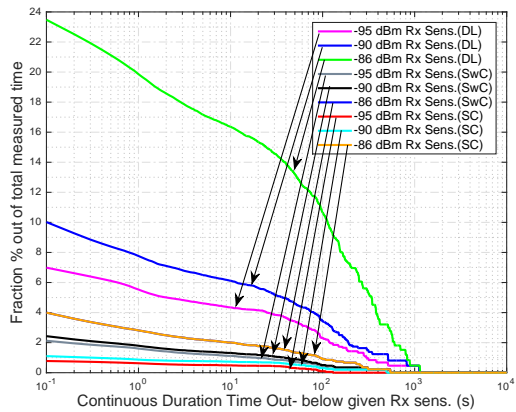}}
\caption{Continuous outage duration below a given $R_x$ sensitivity, with $T_x$  power of 0 dBm, from agglomerate on-body data for direct link (DL), switch-and-examine combining (SwC) and selection combining (SC)}
\end{figure}

\begin{figure}
\centering{\includegraphics[width=\figwidth]{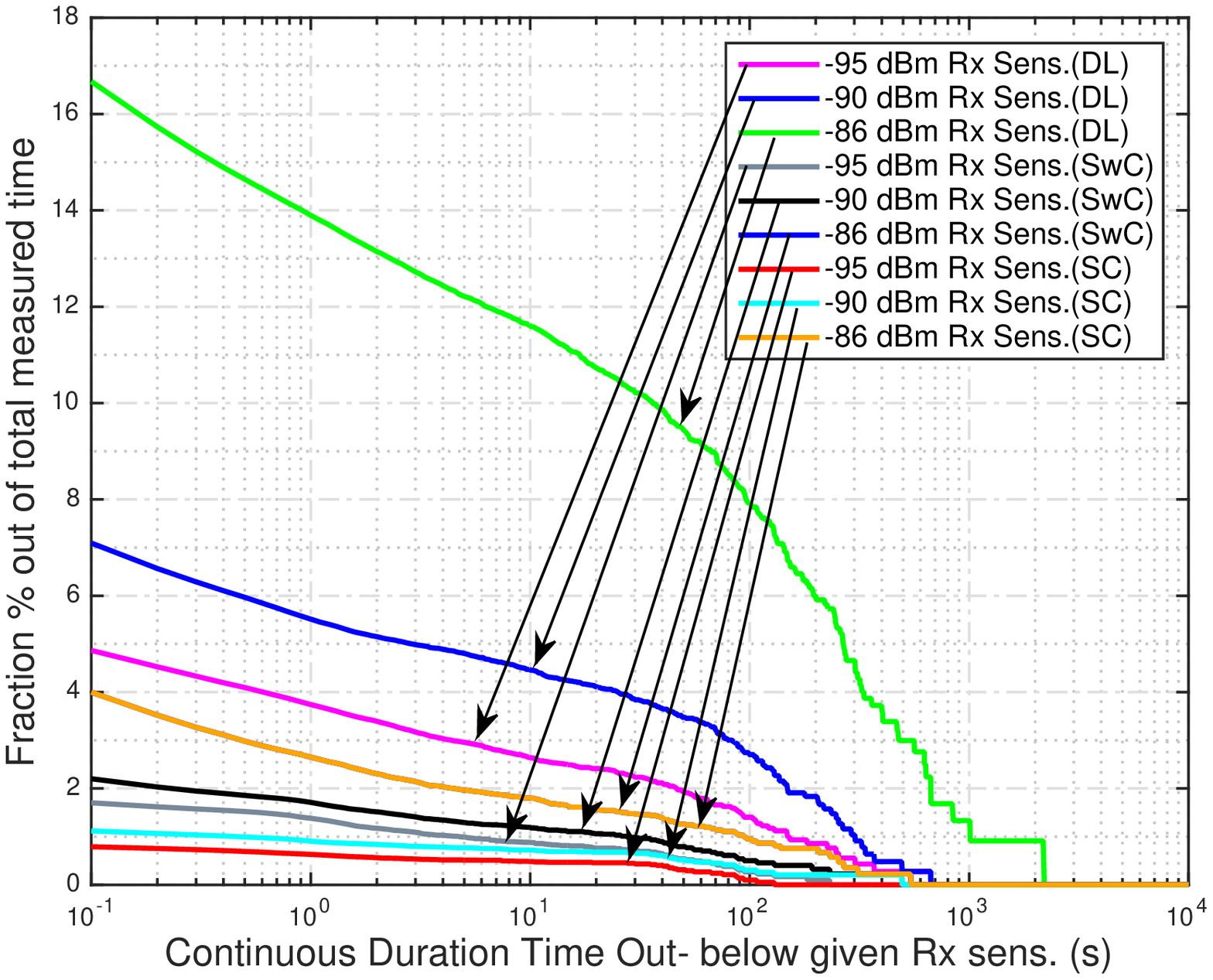}}
\caption{Continuous outage duration below a given $R_x$ sensitivity, with $T_x$  power of 0 dBm, from agglomerate off-body data for direct link (DL), switch-and-examine combining (SwC) and selection combining (SC)}
\end{figure}

\begin{table}[h]
\processtable{Empirical result analysis for direct link (DL) \& cooperatively combined links (SwC \& SC); RS, OP \& COD imply Receive Sensitivity, Outage Probability \& Continuous Outage Duration, respectively}
{\begin{tabular}{ |l|l|l|l||l|l|l| }\hline
& \begin{tabular}{@{}l@{}}DL(on- \\ body)\end{tabular} & \begin{tabular}{@{}l@{}} SwC(on-\\ body)\end{tabular}
& \begin{tabular}{@{}l@{}} SC(on-\\ body)\end{tabular} & \begin{tabular}{@{}l@{}} DL(off-\\body)\end{tabular}
& \begin{tabular}{@{}l@{}} SwC(off-\\body)\end{tabular} & \begin{tabular}{@{}l@{}} SC(off-\\body)\end{tabular}\\\hline
\begin{tabular}{@{}l@{}}Best case\\OP\end{tabular} & $9.1$\% & $3.3$\% & $1.15$\% & $6.3$\% & $3$\% & $1.17$\%\\\hline
\begin{tabular}{@{}l@{}l@{}l@{}}COD\\>10s\\at RS of\\$-86$ dBm\end{tabular} 
& $16.3$\% & $2$\% & $2$\% & $11.5$\% & $1.8$\% & $1.8$\%\\\hline
\begin{tabular}{@{}l@{}l@{}l@{}}COD\\>125ms\\at RS of\\$-86$ dBm\end{tabular} 
& $24$\% & $4$\% & $4$\% & $17$\% & $4$\% & $4$\%\\\hline
\end{tabular}}{}
\end{table}

\section{Conclusion}
We have investigated cooperative communications for an isolated sleeping WBAN with diversity combining techniques. Our results denote that, up to $7$ dB and $20$\% performance improvement can be achieved with cooperative communications, in case of outage probability and outage duration, respectively. It has been shown that, $3$-branch selection combining (SC) provides a slight improvement in diversity gain over $3$-branch switch-and-examine combining (SwC), based on outage probability. In extension of this work, further measurements and analysis will be done with closely located groups of WBANs being used for sleep monitoring.

\vskip3pt
\ack{NICTA is funded by the Australian Government through the Department of Communications and the Australian Research Council through the ICT Centre of Excellence Program.}

\vskip5pt

\noindent S. M. Shimly, S. Movassaghi and D. B. Smith (\textit{The National ICT Australia (NICTA), (CRL/ATP), Australia})
\vskip3pt

\noindent $^*$E-mail: Samiya.Shimly@nicta.com.au
\vskip3pt

\noindent The authors are also with The Australian National University (ANU), Canberra, ACT 0200, Australia

\end{document}